# Monopoly Unveiled: Telecom Breakups in the US and Mexico[1]

Fausto Hernández Trillo*

C. Vladimir Rodríguez-Caballero** [2]

Daniel Ventosa-Santaulària*

## Abstract

This paper posits the decline in market capitalization following a monopoly breakup serves as a means to gauge how financial markets assess market power. Our research, which employs univariate structural time series models to estimate the firm's value without the breakup and juxtapose it with actual post-divestiture values, reveals a staggering drop in AT&T's value by 65% and AMX's by 32% from their pre-breakup levels. These findings underscore the contemporary valuation of monopoly rents as perceived by financial markets, highlighting the significant impact of monopoly breakup on market capitalization and the need for a deeper understanding of these dynamics.

*JEL Classification*: G1, C1, D0

*Keywords*: Monopoly breakup, market power, firm value, ATT, AMX.

[1] We thank Alexander Elbittar and Daniel Ruiz for their helpful comments. Usual disclaimers apply.

[2] Corresponding author. vladimir.rodriguez@itam.mx

* División de Economía, CIDE. Carr. Mx-Toluca 3655. Santa Fe. Mexico City, 01210. Mexico

** Department of Statistics, ITAM. Río Hondo No.1, Col. Progreso Tizapán, Álvaro Obregón, CDMX. 01080. Mexico

## 1. Introduction

At the heart of market power lies a firm's capacity to elevate the price of its product beyond what would be observed in a competitive market setting. Evaluating an industry's proximity to the competitive benchmark involves two widely employed approaches: the rate of return[3] and the price-cost margin.[4] Moreover, in scrutinizing how performance fluctuates with market structure, the literature has consistently endeavored to gauge market power, see Bresnahan (1989). Among the predominant methodologies, one of the most prevalent involves estimating parameters that quantify the degree of competition.

What distinguishes this paper is its innovative exploration of how financial markets assign value to market power. We adapt Faccio's (2006) methodology for estimating political connections, a novel approach in this context. Faccio's work evaluates a company's devaluation resulting from a loss in political connections. Similarly, our study estimates a firm's value decline following a monopoly breakup. To accomplish this, we analyze the stock price performance of prominent cases such as AT&T and the Mexican telecommunications company (formerly a monopoly, Teléfonos de México,[5] hereafter AMX) before and after the monopoly breakup. This loss of corporate value serves as a proxy for understanding how financial markets assess the economic rents derived from market power, presenting a fresh perspective in the field.

The challenge in quantifying the value loss lies in projecting the firm's value without a breakup, enabling a meaningful comparison with the actual value at various points in time.

[3] See Fisher and McGowan (1983) for a review and a critique of this approach.
[4] See Fisher (1987) for a critique on its use.
[5] As we explain later this company had an important market power, in which there was an asymmetry between this firm and other smaller firms competing in or trying to enter to the market. (Alcázar and Pastrana, 2023).

The employed statistical methodology adopts a univariate structural time series model, offering a dynamic representation of market capitalization (referred to as a market cap, calculated as the stock price multiplied by the number of outstanding shares). These models aim to deconstruct the underlying process into permanent components, including level, slope, cycle, and seasonal elements, alongside an irregular or transitory component. We assume an additive integration of these components in the model formulated in state space form (SS). Leveraging the Kalman Filter and Smoothing (KFS) algorithms, the model facilitates maximum likelihood estimation of the unknown parameters within the structural framework.

The results indicate a significant and abrupt decline in both companies' stock prices following the monopoly's dissolution. This observation holds even after accounting for relevant factors. Consequently, the overall loss in value diminishes, assuming a constant debt value, a calculation that has been performed. Specifically, our findings reveal a value loss of approximately 65% for AT&T and 32% for AMX. In essence, these figures underscore the market's perception of the worth of the previously held market power.

The subsequent sections of this paper are organized as follows: Section 2 provides a concise review of the pertinent literature, Section 3 delineates the telecommunications sector, Section 4 introduces the model and discusses the results, and Section 5 presents concluding remarks.

## 2. Literature Review

To address this issue, Thomadakis (1977) proposes using the firm's value, as it mirrors the market's expectation of its ability to sustain excess profits. The underlying rationale is that the capital market assimilates all available information concerning the firm's future profitability, making market value a reflection of the *ex-ante* rate of return on investment.

However, the rates of return approach has its shortcomings. Fisher and McGowan (1983) argue that capital is often improperly valued, with economic definitions neglected in favor of accounting definitions. Moreover, they posit that these rates may not be adequately adjusted for risk. Responding to these challenges, some researchers employ the price-cost margin as an alternative. However, this approach faces difficulties, primarily due to the scarcity of accurate data on marginal costs. Fisher (1987) contends that such measures are unreliable and plagued by capital valuation issues.

In essence, the debate on this matter persists. A breakup, indicative of heightened competition, may stimulate innovation, see Miller (1995). However, the outcome remains uncertain, and a decline in stock prices might be anticipated in the short run.[6] This paper seeks to estimate the value of market power using the firm's market value. The subsequent section elucidates our approach in detail.

## 3. The Monopoly Power of AMX AND AT&T

Founded in the late 1800s, AT&T, a telecommunications company, was historically viewed as a regulated monopoly. In the 1970s, concerns arose regarding potential violations of the Sherman Antitrust Law. During this period, AT&T retained control over long-distance services. However, in 1984, as part of an agreement with the authorities, the local service segment underwent a significant restructuring, creating seven "baby-bells."

The breakup brought about several benefits for consumers. Customers were free to purchase phones manufactured by other companies, leading to a subsequent price drop. Competition

[6] In contrast, the stock market response to mergers is positive for the combined merging entities, suggesting that mergers create shareholder value (see Andrade et al., 2001; Hackbarth and Morellec, 2008).

in the long-distance service sector also intensified, resulting in lower charges, see Temin (1987).

A few years after the breakup in the 1990s, AT&T embraced internet technology, transforming itself into an internet service provider. Today, the company offers diverse services, aligning with its commitment to various connection forms, including digital and wireless services.

In 1990, Telmex (AMX), the state-owned telephone company in Mexico, underwent privatization, emerging as the sole player in the fixed-line service market. By 1998, it boasted a staggering 100% market share in fixed-line telephony, which only slightly diminished to over 80% by 2010. As the telecommunications landscape evolved, the company strategically diversified by venturing into cellular phone services, eventually claiming a formidable 79.6% market share, see OECD (2011).

However, despite its dominance, pricing for both fixed-line and cellular services exceeded international benchmarks. The excessive pricing of telecommunication services in Mexico from 2005 to 2009 led to an estimated average annual loss in consumer welfare equivalent to 1.8% of Mexican GDP, see OECD (2011).

The company wielded substantial market power. In response to mounting pressure from civil society, the Mexican government implemented industry breakup measures through asymmetric regulation to foster increased competition (Alcázar & Ramos (2023)). The subsequent result was a shift to more competitive pricing.

The financial implications of this breakup pose an intriguing question: What was the corporate value loss incurred as a result? This loss approximates the market power's value—essentially, the financial market's assessment of AMX's monopoly-like position.

Figure 1 illustrates the Securities Market Line (SML) for the Mexican stock market from 2000 to 2008. It is crucial to note that this timeframe represents the pinnacle of AMX's market power, as indicated by the highlighted circle in the graph. Positioned well above the SML, AMX notably dominates the majority of firms stochastically. By 2013, AMX had evolved into a conglomerate formed by twin companies Telmex and Amovil, consistently surpassing the SML.

Remarkably, even when factoring in risk considerations, AMX exhibits an excess return, suggesting a valuation beyond expected based on risk alone. This excess return can be interpreted as the market's assessment of the firm's market power. In 2014, recognizing the need for increased competition, the Mexican Congress approved a telecommunications reform to break up AMX's monopoly by facilitating the entry of competitors, with AT&T being among them.

**Figure 1:** Securities Market Line for Mexico

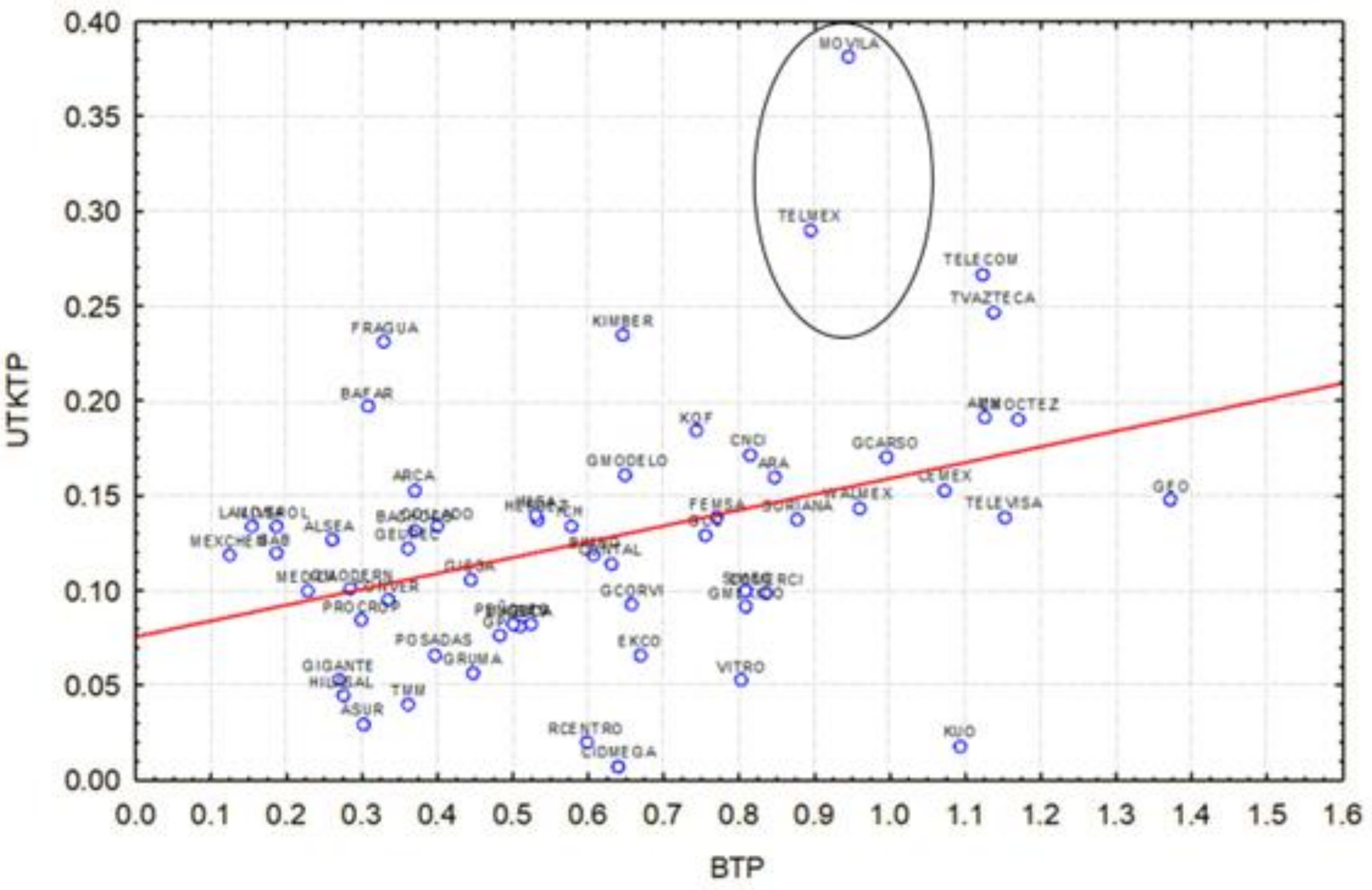

Therefore, there is no necessity to calculate an appropriate rate of return or a price-cost margin to discern the cost structure for deducing the present value of monopoly profits as perceived by markets. Subsequently, we proceed to estimate this loss.

## 4. Methodology

Faccio (2006) argues that financial markets assign a higher value to firms with political connections; hence, when these are lost (for example, after Suharto's death in Indonesia), the value of stocks goes down abruptly. The author interprets this loss of value as the value the market assigns to political connections.

We follow this rationale. We argue that when a monopoly breakup occurs, the value of the monopolist firm goes down. This amount may be interpreted, *ceteris paribus*, as the present value of the future excess profits. In other words, the loss of value may be interpreted as how much investors value the monopoly power over time.

Our methodology involves projecting the firm's value without a breakup to establish the baseline value. Subsequently, we compare this projection to the actual value at each point in time. The ensuing section outlines the statistical methodology employed in detail.

We specify a univariate structural time series model to describe the market capitalization (stock price times the number of outstanding shares, market cap hereafter) behavior as a dynamic system. These models decompose the underlying process into permanent components (level, slope, cycle, and seasonal) and an irregular or transitory part. We assume that components enter into the model in an additive fashion. The model is written in state space form (SS, hereafter) to employ Kalman Filter and Smoothing (KFS) algorithms to open the way to the maximum likelihood estimation of the unknown parameters of the structural model.

In the model, the trend component is the time series level, which fluctuates up and down according to the underlying long-term tendency. The slope component accounts for short-term fluctuations around the trend, the cycle component represents periodic or recurrent patterns, and the seasonal component captures regular seasonal variations. Such a model is specified as follows. We first introduce the output equation, which relates the system's internal structure to the observed variable:

$$y_t = \mu_t + \gamma_t + c_t + \varepsilon_t, \qquad \varepsilon_t \sim N(0, \sigma_\epsilon^2), \tag{1}$$

and then include the state equations that describe the internal dynamics and, precisely, their state from one-time step to the next:

$$\mu_{t+1} = \mu_t + \beta_t + \xi_t, \qquad \xi_t \sim N(0, \sigma_\xi^2), \tag{2}$$

$$\beta_{t+1} = \beta_t + \zeta_t, \qquad \zeta_t \sim N(0, \sigma_\zeta^2), \tag{3}$$

where $y_t$ is the observed variable, $\mu_t$ is the trend component (it includes level and slope), $\gamma_t$ is the seasonal component, $c_t$ is the cycle component. We define the seasonal component with period $s = 12$ with the following trigonometric specification:[7]

$$\gamma_t = \sum_{j=1}^{[s/2]} \gamma_{j,t},$$

$$\gamma_{j,t+1} = \gamma_{j,t} \cos(\lambda_j) + \gamma_{j,t}^* \sin(\lambda_j) + \omega_{j,t}, \tag{4}$$

$$\gamma_{j,t+1}^* = -\gamma_{j,t} \sin(\lambda_j) + \gamma_{j,t}^* \cos(\lambda_j) + \omega_{j,t}^*, \tag{5}$$

for $j = 1,2, \ldots, [s/2]$, where $\omega_{j,t}$ and $\omega_{j,t}^*$ are two mutually uncorrelated white noise disturbances with zero means and common variance $\sigma_{\omega_j}^2$, and $\lambda_t = 2\pi j/s$ is the frequency in radians.

As for the cycle component with period s:

$$c_{t+1} = c_t \cos(\lambda_c) + c_t^* \sin(\lambda_c) + \kappa_t, \tag{6}$$

$$c_{t+1}^* = -c_t \sin(\lambda_c) + c_t^* \cos(\lambda_c) + \kappa_t^*, \tag{7}$$

[7] To simplify the computation of the seasonal component without a significant adjustment loss, we follow Hindrayanto, Aston, Koopman & Ooms (2013) by assuming only two different frequencies. This allow us to reduce the number of parameters to be estimated. The problem lies in the fact that, when all frequencies are considered, the optimization of the loglikelihood is inefficient. For further details, see Hindrayanto, Aston, Koopman & Ooms (2013).

where $\omega_t$ and $\omega_t^*$ are, again, $iid\ N\left(0, \sigma_{\kappa_j}^2\right)$, and $\lambda_c = 2\pi/s$.[8] All disturbances involved in equations 1-5 are mutually independent.

Together, the state equations and the output equations form the structural model.

To specify the SS model, we must consider several factors: the characteristics of the data (visual inspection), the system's underlying dynamics, and the model's objective. The goal is to retain the main properties of the dynamics of the market cap series and reproduce them faithfully if such dynamics have stayed the same; this ensures a credible counterfactual estimation of the price series had the monopoly not been broken up.

We prioritize testing the components' variances to decide what features to include in each case. The latter procedure is essential because of (i) identifiability (the variances of the different components play a crucial role in determining the behavior and characteristics of the model. If the variances are set to zero, it implies that the corresponding components have no variability or uncertainty; non-zero variances ensure that there is sufficient variability in the components, enabling better parameter estimation and avoiding degenerate solutions); (ii) numerical stability (non-zero variances provide numerical stability and facilitate the estimation process); and (iii) adequateness of the specification (when a variance component is estimated to be zero, it implies that the corresponding component is not contributing to the variation in the data. This indicates that the component is irrelevant or the model formulation is inadequate).

[8] The period of the cycle component is *s=96* months, this is, eight years.

It is relevant to clarify the possible SS specifications from equations 1-7:

- When $\sigma_{\xi}^{2} = 0$ and $\sigma_{\zeta}^{2} \neq 0$, the SS specification is called Integrated Random Walk. Note that $y_t \sim I(2)$ .
- When $\sigma_{\xi}^{2} \neq 0$ and $\sigma_{\zeta}^{2} = 0$, the SS specification is a Random Walk plus drift**.** Note further that when $\beta_t = 0$, the SS becomes a simple Random Walk. Note that in this case, $y_t \sim I(1)$, but there is also a deterministic trend that dominates the former I(1) case.
- When $\sigma_{\xi}^{2} \neq 0$ and $\sigma_{\zeta}^{2} \neq 0$, the SS specification is called Local Level Trend. Note that in this case, $y_t \sim I(2)$. That said, there is a mixture of two integrated processes: An $I(2)$ process (from $\beta_{t+1}$) and an $I(1)$ process (from $\mu_{t+1}$). Nonetheless, the $I(2)$ process dominates the dynamics over the $I(1)$ process.
- When $\sigma_{\xi}^{2} = 0$ and $\sigma_{\zeta}^{2} = 0$, the SS specification is a Local Level. Note that $y_t \sim I(0)$. Also note that when $\beta t \neq 0$, a linear trend surges, and $yt$ is Trend stationery. Otherwise, when $\beta_t = 0$, $y_t$ is *stationary*.

Two SS specifications are relevant for our purposes: the Integrated Random Walk and the Random Walk plus Drift. We test the null hypotheses: a) H0: $\sigma_{\zeta}^{2} = 0$ and b) H0: $\sigma_{\xi}^{2} = 0$ to distinguish between both cases. Furthermore, a stochastic seasonal component at a particular frequency is found when $\sigma_{\omega_j}^{2} > 0$. Similarly, the stochastic cyclical component is granted when $\sigma_{\kappa_j}^{2} > 0$.

In short, we proceeded as follows:

We first follow Harvey (2001) to test whether there is an integrated random walk plus noise model ($H_0: \sigma_\xi^2 = 0 \; vs \;\; Ha: \sigma_\xi^2 > 0$). Secondly, once we get evidence of the first step, we test whether the trend is reduced to a random walk with drift, that is $H0: \sigma_\zeta^2 = 0 \; vs \;\; Ha: \sigma_\zeta^2 > 0$.[9]

To test whether the seasonal component is stochastic, we follow Canova and Hansen (1995), Franzini and Harvey (1983), and Busetti and Harvey (2003).

For the cycle component, we adopt a different approach. First, we employ a "sample size" feasibility criterion. For instance, the cycle is not included in the case of AMX because the selected period is insufficiently long to be credibly estimated (8 years, only one cycle). Second, for the case of AT&T, we compare the information criteria to decide whether the cycle is worth including.

Our dataset[10] spans 1971 to 1992 for AT&T and 2011 to 2017 for AMX. However, to ensure that SS models precisely capture the dynamics of each series before the breakup, we restricted the samples to the periods 1971 to 1983 (AT&T) and 2011 to July 2015 (AMX). Data frequency is monthly.[11]

It is essential to highlight that neither of the samples incorporates data from the post-breakup period. This choice holds particular significance, as our objective is to replicate the dynamics of the price series that characterized the period preceding the breakup. With the structural time series (SS) model established, we leverage it to construct the counterfactual for that specific timeframe.

---

[9] Under the null, the test statistic has a Cramér-von Mises law. See also Harvey (2001) for further details.
[10] Data for AT&T was obtained at https://investors.att.com/stock-information/historical-stock-information/historical-quote/att-corp ("Stock Information," Quote, AT&T Corp, April first, 2023). As for AMX data was obtained at https://es.finance.yahoo.com/quote/AMX/history?p=AMX&guccounter=1 (América Móvil, S.A.B. de C.V. (AMX), April first, 2023).
[11] Original data is daily. Nonetheless, the number of missing observations was substantial. Missing data could have been imputed but we opted to reduce the frequency by computing month averages. This comes at a price of reduced variability, but the phenomenon under study (value loss because of a breakup) has little to do with daily variability.

The counterfactual inherently assumes that conditions remained unchanged, offering an estimate of the price trajectory had the breakup not occurred. This approach, referred to as the counterfactual strategy, allows us to examine how the price series would have evolved without the breakup.[12]

### 4.1 Results

The findings can be briefly summarized as follows. In both instances, the evidence strongly indicates that a random walk with drift and stochastic seasonality is the appropriate model to capture the time-series dynamics of the respective series accurately. As anticipated, the ATT case incorporated the cycle component in the specification, while this was not the case for AMX.[13] The output specifications are as follows:

- AT&T: $p_t = \mu_t + \gamma_t + c_t + \varepsilon_t,$
- AMX: $p_t = \mu_t + \gamma_t + \varepsilon_t,$

where $p_t$ is the market cap, either AT&T or AMX. As for the State equations, we set:

- $\mu_{t+1} = C + \mu_t + \xi_t,$ (2)

where C is a constant term.

---

[12] SS models are estimated using Quasi-Maximum Likelihood and KFS algorithms, see Harvey (1990) for technical details. We use the software R to estimate all models. Particularly, we employ the KFAS toolbox developed by Helske (2017).

[13] The data set is not long enough to encompass a reasonable number of cycles.

The seasonal component for AT&T and AMX is written as in equations (4) & (5), considering the specific statistically significant frequencies. The cycle component (only for AT&T) is specified as in equations (6) & (7). Table 1 presents the results:

**Table 1:** Variance tests on the SS models for AT&T and AMX

| Testing against a stochastic level Ho: $\sigma_{\xi}^{2}=0$ | | | | |
|---|---|---|---|---|
| AT&T | AMX | Degrees of freedom | | |
| 1.022*** | 3.264*** | 1 | | |
| Testing against a stochastic level Ho: $\sigma_{\zeta}^{2}=0$ | | | | |
| AT&T | AMX | Degrees of freedom | | |
| 0.019 | 0.112 | 1 | | |
| Testing against a stochastic level Ho: $\sigma_{\omega}^{2}=0$ | | | | |
| Frequency | | AT&T | AMX | Degrees of freedom |
| Individual frequencies | Group: $\omega_1$ | 0.521 | 0.353 | 2 |
| | | 3.116*** | 1.076*** | 2 |
| | | 0.665* | 0.291 | 2 |
| | | 0.871** | 0.517 | 2 |
| | | 0.457 | 1.252*** | 2 |
| | | 0.583** | 0.441* | 1 |
| | Group: $\omega_2$-$\omega_6$ | 5.694*** | 3.579*** | 9 |
| All | | 6.214*** | 3.933*** | 11 |

Level test: Harvey & Streibel (1997) test for the presence of a random walk component. Seasonality tests: Canova and Hansen (1995) show that the asymptotic distribution of this statistic is generalized Cramer-von Mises with the degrees of freedom in the last column. A joint test against the presence of stochastic trigonometric components at all seasonal frequencies is based on a statistic obtained by summing the individual test statistics. This statistic has an asymptotic distribution, generalized Cramer-von Mises with s-1 degrees of freedom, see Busetti & Harvey (2003).

Importantly, we conduct diagnostic checks on the model (independence, homoskedasticity, and normality) to assess if the model's parameters are efficiently estimated using the maximum likelihood approach at the estimation stage. Additionally, we present the log-likelihood and compute three standard information criteria (AIC, BIC, and AICc). The results of these tests are displayed in Table (2).

**Table 2:** Diagnostic tests of SS models for AT&T and AMX

| Company | AT&T | | | AMX | |
|---|---|---|---|---|---|
| Test/criterion | BLM | No cycle | No cycle, only 2 seasonal frequencies | BLM | Only 2 seasonal frequencies |
| Ljung–Box[24] test | 20.3** | 18.73* | 26.48 | 19.359* | 19.369* |
| Heteroskedasticity | 0.75 | 0.78 | 0.79 | 0.901 | 0.905 |
| LV Test (Normality) | 0.69 | 0.91 | 1.89 | 1.701 | 1.665 |
| Log-Likelihood | 275.82 | 278.57 | 276.24 | 50.774 | 50.771 |
| AIC | -521.64 | -531.14 | -542.49 | -75.548 | -75.543 |
| BIC | -474.78 | -490.53 | -526.87 | -47.083 | -47.077 |
| AICc | -518.48 | -528.78 | -542.12 | -68.548 | -68.543 |

Ljung-Box and Heteroskedasticity tests: lags in brackets, see Harvey (1990). LV test of Lobato & Velasco (2004).

As evident from the results presented in Table (2), there is some indication of autocorrelation in both cases (AT&T and AMX), while no evidence of heteroskedasticity is observed. Furthermore, the residuals exhibit strong evidence of normality. Additional testing procedures detailed in Appendix A further support the residuals' normality, independence, and stability.[14] Moreover, the information criteria in Table (2) suggest the following:

- For AT&T: Including the cycle component enhances the model's fit to the data.

[14] In appendix A the reader can find additional tests to ensure the adequateness of the proposed specifications for both AT&T and AMX: the standardized residuals, the QQ plot (also, of residuals), the Ljung-Box (autocorrelation) test statistic, the sample autocorrelation function of standardized residuals, and the CUSUM plot to verify stability, again, of residuals.

- For AT&T and AMX: Allowing for all frequencies in the seasonal component leads to improved model performance.

### 4.2 Counterfactual building

After estimating the structural models, we leverage the results to construct the counterfactual behavior of market capitalization prices. The forecasted periods extend from January 1984 to June 1989 for AT&T and from July 2015 to November 2016 for AMX. The rationale behind defining these subsamples is twofold: the starting points align with the data of the judicial monopoly breakup decision, and the endpoints correspond to the months where the counterfactual becomes statistically indistinguishable from the observed market capitalization.[15]

Figures (2) and (3) showcase the outcomes of the AT&T and AMX breakups, respectively. These illustrations depict the evolution of each company's market capitalization before and after the event, alongside the counterfactual derived from the pre-breakup period using the methodology described earlier.

Monopoly breakups often evoke uncertainty, leading numerous investors to reduce their stock holdings or exit the market altogether. Share prices in both cases declined noticeably, with a 1-year average price drop[16] of 74% for AT&T and 32% for AMX. However, the recovery process for both firms takes time as they navigate innovative strategies in a more competitive environment post-breakup.

---

[15] In the AMX case, the counterfactual becomes significantly different from the observed series for a short period (July-15, August-15 and October-15 for AMX and July-) just at the start of the breakup. We nonetheless include those months in the counterfactual sample. In the AT&T case, it becomes significantly different from the observed series after it had already become non-significantly different (August-90 – March-91). In this case we omitted this period in the counterfactual sample. Being stricter with the definition of the subsamples does not compromise our results.

[16] Average Market Cap of previous year over the one of the post breakup.

Figure 2 shows that AT&T's stock price "reaches into" the confidence interval five years after the breakup. Consequently, the market capitalization loss is estimated as the present value of these five years. The estimate is presented in Table 3 and amounts to 65%.

**Figure 2:** AT&T Market Cap (in logs). Observed vs counterfactual.

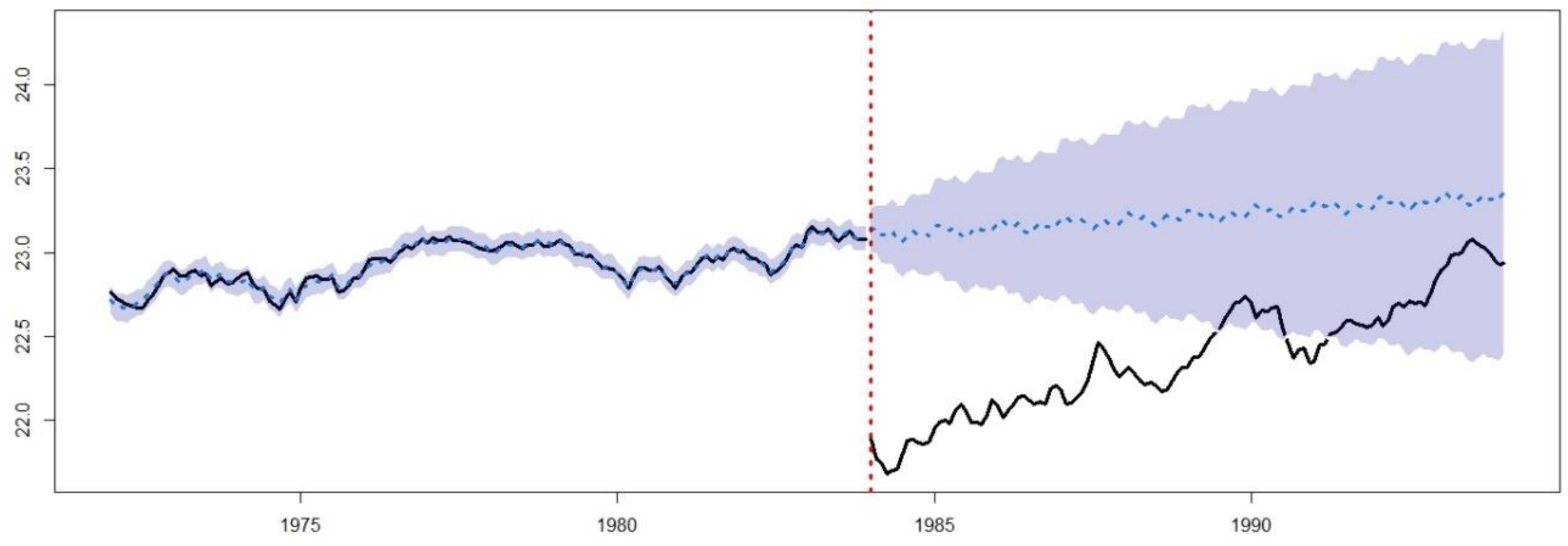

**Table 3. ATT Loss of Value after Break up**

| | *Loss of Value (millions of USD)** | *% Loss of Value** |
|---|---|---|
| 1-yr Average after breakup | -7,900,718,045.98 | -72.9% |
| 2-Yr Average from after breakup | -7,678,369,840.63 | -70.9% |
| 3-yr Average after breakup | -7,573,666,849.27 | -69.9% |
| **5-yr-Average** after breakup | **-7,048,522,354.60** | **-65.0%** |

* The difference between forecasted and actual stock price times the number of shares in circulation

For the AMX case, Figure 3 shows that its stock price moves into the confidence interval after two years. The present value of the market cap loss is nearly 32% of the company's value (assuming an unlevered position), as presented in Table 4.

**Figure 3:** AMX Market Cap (in logs). Observed vs counterfactual.

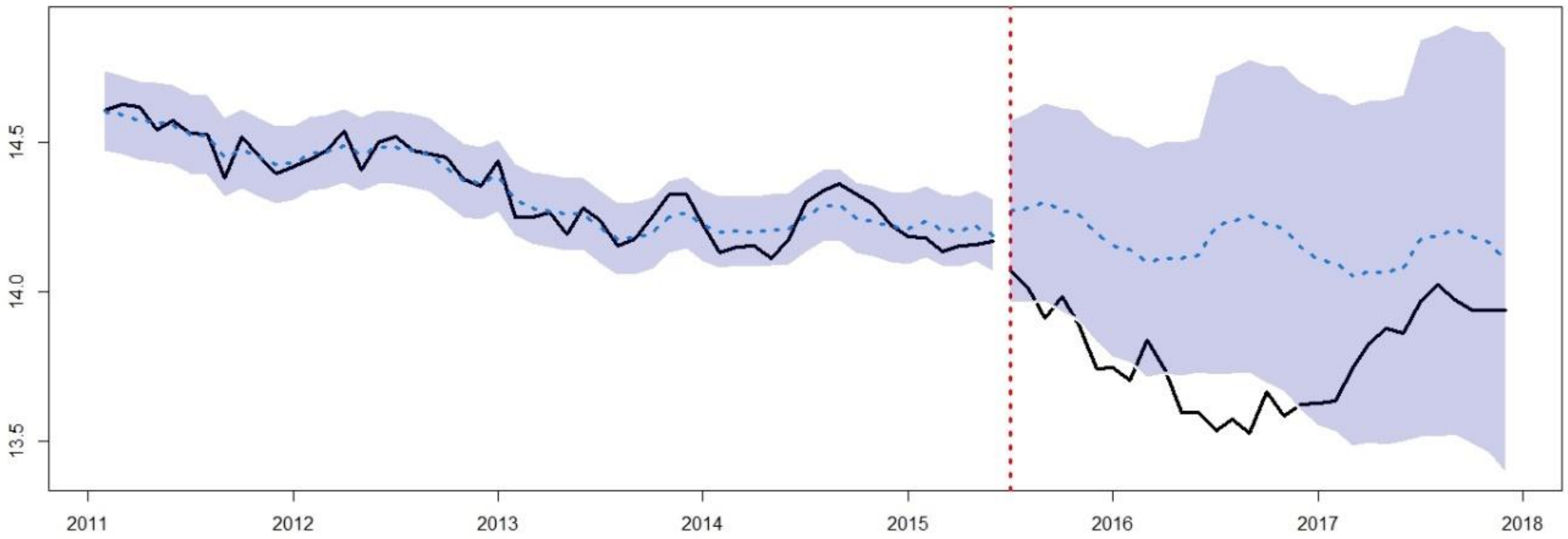

**Table 4. AMX Loss of Value after Break up**

| | *Loss of Value (millions of USD)** | *% Loss of Value** |
|---|---|---|
| 1-yr Average after breakup | -447,295.72 | -29.29% |
| 2-Yr Average from after breakup | -485,531.21 | -31.80% |
| 3-yr Average after breakup | -441,970.10 | -28.95% |

* The difference between forecasted and actual stock price times the number of shares in circulation

We contend that the loss of value (assuming an unlevered company) signifies how financial markets assessed the extraordinary rents derived from the monopolist position.

Conventional theory posits that competitive markets enhance efficiency and stimulate innovation, implying that a monopoly breakup could eventually lead to a recovery in financial markets (Miller, 1995). The duration of this restorative process remains uncertain. However, as previously mentioned, we posit that it occurs when the firm realigns with the stochastic process's trajectory (i.e., within the confidence interval).

A notable aspect is the disparity in the value loss between the two firms: AT&T experienced a 65% loss, while AMX's loss was 32%, half of the former. Furthermore, the recovery speed for AMX is also half the time it took for AT&T. Mexico's weak democracy and the rule of law add a layer of complexity to this scenario. The country ranks 89th out of 132 in the Democracy Index by the Economist Intelligence Unit[17] and 113th out of 134 in the World Justice Project's Rule of Law Index. [18] Acemoglu and Robinson's book "Why Nations Fail" (2012) highlights the nation's weak institutions. Consequently, we argue that the impact of the breakups varied.

Despite this, there has been increased competition within the telecommunications sector in Mexico. Alcazar and Ramos (2023) prove that asymmetric regulations promote competition by facilitating sector entry. Simultaneously, these regulations intensified competition by enhancing the quality of services offered by the preponderant agent (AMX), resulting in a net gain in the number of transferred lines and an increase in postpaid lines.

---

[17] https://www.eiu.com/n/wp-content/uploads/2023/02/Democracy-Index-2022_FV2.pdf?li_fat_id=f1fbad7e-a282-4b9e-9f8f-6a6d5a9fe6b8 (consulted on November 16, 2023)
[18] (consulted on November 16, 2023)

In summary, the market cap loss can be interpreted as a reflection of how financial markets value market power. It is crucial to underscore that this does not represent the actual monopoly profits but reflects how financial markets assess those profits through fundamental analysis.

## 5. Final Remarks

This paper endeavored to estimate the corresponding value loss in the United States and Mexican contexts to assess the economic repercussions of a monopoly breakup within the telecommunications industry. The focal point of our interpretation lies in discerning how financial markets appraise market power. Our methodology involves forecasting the market capitalization for two critical periods: January 1984 to June 1989 for AT&T and July 2015 to November 2016 for AMX. The definition of these subsamples aligns with a strategic rationale: the starting points correspond to the dates of judicial monopoly breakup decisions, while the endpoints signify the months where the counterfactual becomes statistically indistinguishable from the observed market capitalization. Subsequently, we compare these projected values to the actual market observations.

The outcomes reveal a stark contrast in the value loss experienced by the two firms. AT&T incurred a substantial 70% reduction compared to its pre-breakup value, whereas AMX faced a more moderate figure of 30% of its original value. This discrepancy suggests that the market may perceive these reductions as reflective of the present value of monopoly rents.

The implications of these findings extend beyond the immediate market dynamics. The substantial disparity in the value loss between the two firms raises intriguing questions about the factors influencing market reactions to monopoly breakups. This discrepancy prompts a deeper exploration into the nuanced market conditions, institutional frameworks, and regulatory environments that shape the aftermath of such transformative events.

In conclusion, this paper contributes to our understanding of the economic consequences of monopoly breakups by shedding light on how financial markets evaluate the inherent market power. The divergence in value losses observed between AT&T and AMX underscores the complexity of market dynamics. It provides a foundation for further research into the intricacies of market responses to monopoly transformations in the telecommunications sector.

# Appendix A: Robustness checks

## A.1: diagnostic checks on the AT&T model.

Figure A.1 shows several tests on the AT&T residuals: (a) the standardized residuals, (b) the QQ plot, (c) the Ljung-Box (autocorrelation) test statistic, (d) the sample autocorrelation function (of standardized residuals), and (e) the CUSUM plot to verify stability.

**Figure A.1:** Extended diagnostic checks for AT&T Stock Price.

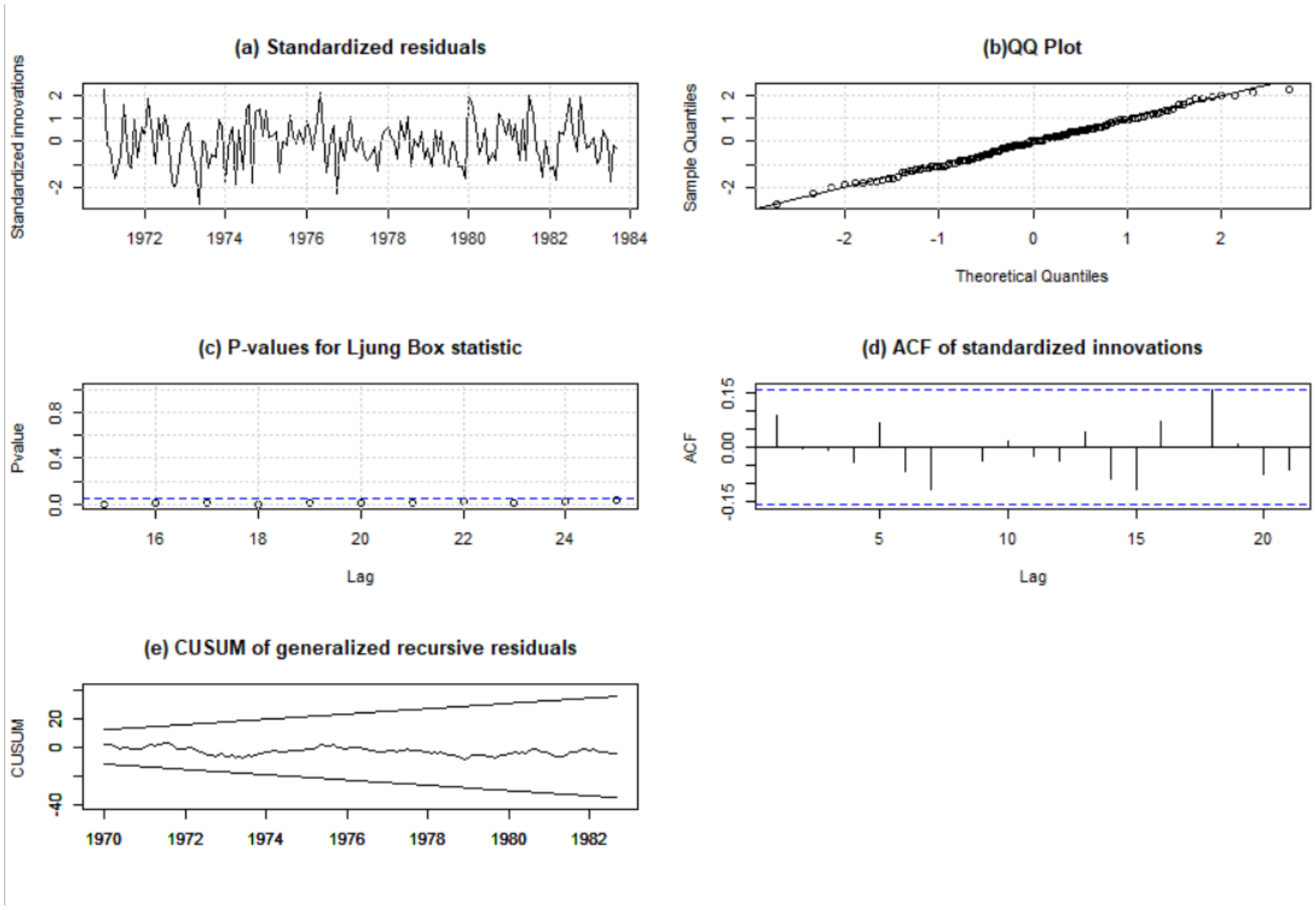

As can be seen from the results exhibited in Figure A.1, the SS specification for AT&T can be considered normally distributed, independent, and stable.

## A.2: diagnostic checks on the AMX model.

Figure A.2 shows several tests on the AMX residuals: (a) the standardized residuals, (b) the QQ plot, (c) the Ljung-Box (autocorrelation) test statistic, (d) the sample autocorrelation function (of standardized residuals), and (e) the CUSUM plot to verify stability.

**Figure A.2:** Extended diagnostic checks for AMX Stock Price.

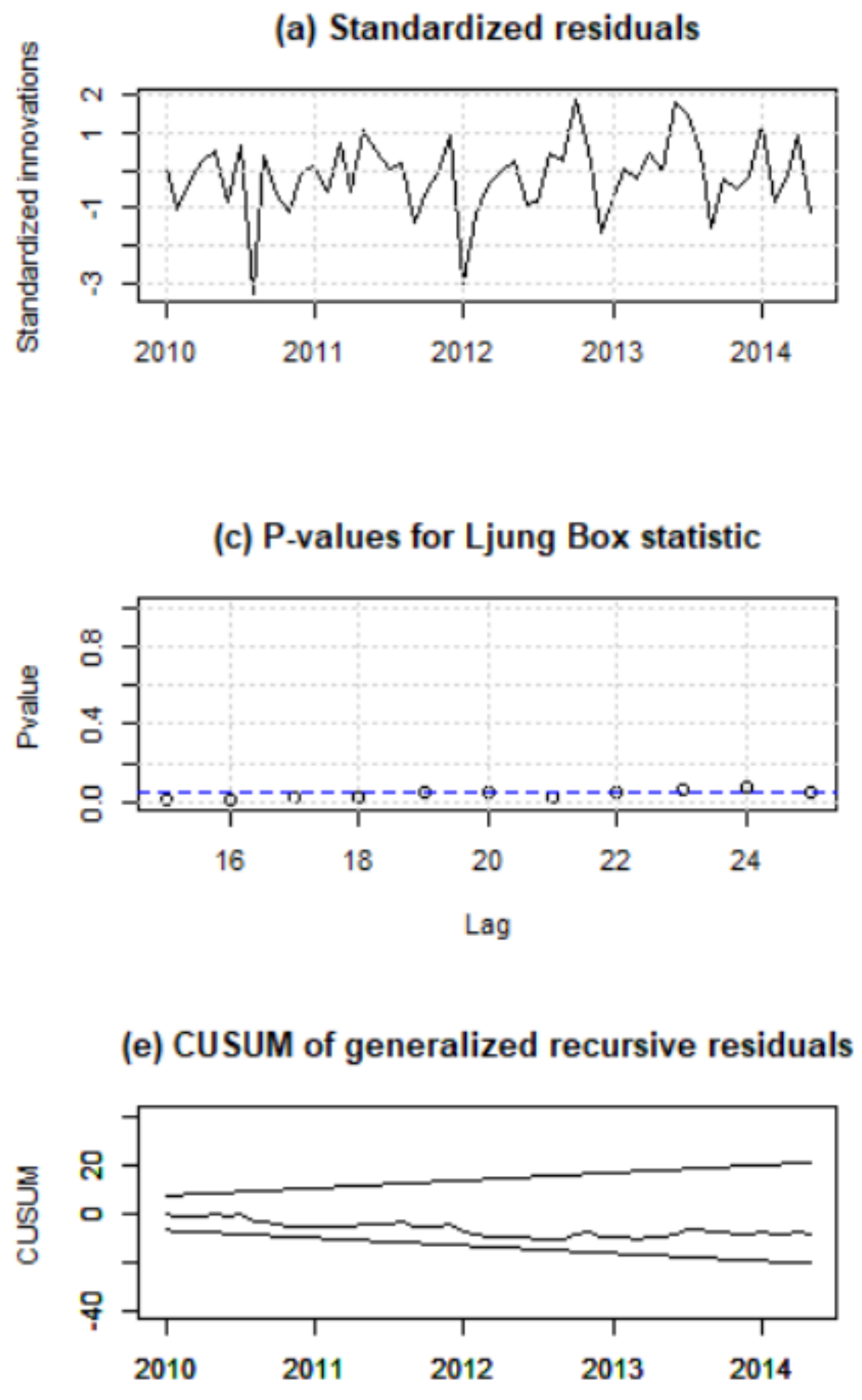

As can be seen from the results exhibited in Figure A.2, the SS specification for AMX can be considered normally distributed, independent, and stable.